\documentclass[reprint,
superscriptaddress,
 amsmath,amssymb,
 aps,
pra,]{revtex4-1}
\usepackage{blindtext}
\usepackage{float}
\usepackage{graphicx}
\usepackage{subfigure}

\begin{document}
\title{Frustrated double ionization of atoms in circularly polarized laser fields}
%\title{Intensity dependent strong field frustrated double ionization of argon atoms}
\author{HuiPeng Kang}
\email{H.Kang@gsi.de}
\affiliation{Institute of Optics and Quantum Electronics, Friedrich Schiller University Jena, Max-Wien-Platz 1,07743 Jena, Germany}
\affiliation{Helmholtz Institut Jena, Frötbelstieg 3, 07743 Jena, Germany}
\affiliation{State Key Laboratory of Magnetic
Resonance and Atomic and Molecular Physics, Wuhan Institute of
Physics and Mathematics, Innovation Academy for Precision Measurement Science and Technology, Chinese Academy of Sciences, Wuhan 430071, China}

\author{Shi Chen}
\affiliation{School of Physics, Peking University, Beijing 100871, China}
\affiliation{Center for Applied Physics and Technology, HEDPS, and College of Engineering, Peking University, Beijing 100871, China}

\author{Jing Chen}
\affiliation{Institute of Applied Physics and Computational Mathematics, P.O. Box 8009, Beijing 100088, China}
\affiliation{Center for Advanced Material Diagnostic Technology, College of Engineering Physics, Shenzhen Technology University, Shenzhen 518118, China}
\author{Gerhard G. Paulus}
\affiliation{Institute of Optics and Quantum Electronics, Friedrich Schiller University Jena, Max-Wien-Platz 1,07743 Jena, Germany}
\affiliation{Helmholtz Institut Jena, Frötbelstieg 3, 07743 Jena, Germany}
%\date{\today}

\begin{abstract}
We theoretically study frustrated double ionization (FDI) of atoms subjected to intense circularly polarized laser pulses using a three-dimensional classical model. We find a novel ``knee'' structure of FDI probability as a function of intensity, which is similar to the intensity dependence of nonsequential double ionization probability. The observation of FDI is more favourable when using targets with low ionization potentials and short driving laser wavelengths. This is attributed to the crucial role of recollision therein, which can be experimentally inferred from the photoelectron momentum distribution generated by FDI. This work provides novel physical insights into FDI dynamics with circular polarization. 
\end{abstract}

\maketitle

\section{Introduction}
When interacting with an intense laser pulse, the electrons in atoms or molecules can be strongly driven by the laser field, leading to various highly nonlinear phenomena mostly accompanied by photoionization. Interestingly, even for such strong laser field a substantial portion of electrons can be trapped into high-lying Rydberg states rather than being released into the continuum \cite{NubbemeyerPRL2008}. Rydberg state excitation of atoms and molecules has a wide range of applications in acceleration of neutral particles \cite{Eichmann2009nature}, precision measurements \cite{Nature535-262}, generation of near-threshold harmonics \cite{PRL112-233001}, and quantum information \cite{RevModPhys82-2313}. In the past few years there is a growing interest in understanding the mechanism of Rydberg state excitation in strong laser fields \cite{PhysRevLett.110.023001,PhysRevA.87.033415,PhysRevA.89.023421,popruzhenko2017quantum,PhysRevLett.114.123003}.
%mostly with photoionization as the precursor.

The formation of excited Rydberg states has been experimentally identified for rare gas atoms, which is dubbed frustrated tunneling ionization \cite{NubbemeyerPRL2008}. Subsequently it was shown that during molecular double ionization, one of the two emitted electrons may be trapped by the fragments generated by Coulomb explosion of molecules, giving rise to the formation of highly excited neutral fragments and singly charged ion fragments \cite{PRL102.113002}. This process can be coined as frustrated double ionization (FDI). FDI has been the focus of intense studies for various molecular systems such as H$_{2}$ \cite{PRL102.113002,PhysRevLett.119.253202}, D$_{2}$ \cite{PhysRevA.84.043425,PhysRevA.98.013419}, O$_{2}$ \cite{PhysRevA.100.063413}, N$_{2}$ \cite{nubbemeyer2009excited}, CO \cite{PRA101.033401}, and clusters \cite{PhysRevA.82.013412,PhysRevA.82.013413,PhysRevLett.107.043003,PhysRevA.88.065401}. However, FDI of atoms has attracted much less attention. From experimental point of view, molecular FDI can be identified by measuring the kinetic energy release during Coulomb explosion of molecules, which can not be applied to atomic targets without dissociation. Very recently, Larimian \textit{et al.} reported coincident measurements on FDI of Ar atoms \cite{2020LarimianPRR}. It was found that atomic FDI shows a strong dependence on laser intensity. For high intensities where sequential double ionization (SDI) dominates, electron trapping mainly happens during the second ionization step. For modest intensities where nonsequential (recollision-induced) double ionization (NSDI) dominates, the electron momentum distributions produced by FDI show features similar to that from double ionization (DI). Detailed theoretical studies have shown that recollision and ionization-exit velocity distribution play important roles in atomic FDI \cite{2009ShomskyPRA, 2020WangxuOE,ourFDIpaper}.  

Most of the investigations of FDI mentioned above adopted linearly polarized light. It is well known that in the case of elliptical or circular polarization, the additional transverse electric field of the laser light steers away the recolliding electrons and thus reduces the chance of recollision. Therefore, NSDI yields are expected to be decreased rapidly as the light ellipticity is increased, which has been verified for rare gases \cite{PRA50.R3585}. However, significant NSDI contributions have been observed for Mg atoms with circularly polarized light \cite{PRA64.043413}. Theoretical studies employing semiclassical and fully classical models have shown that efficient recollision-induced double ionization with circular polarization is still possible via specific trajectories \cite{PRL105.083001,PRL105.083002,PRL108.103601, PRL110.253002}. On the other side, whether FDI exists and how to understand its dynamics in circularly polarized laser fields remains an open question.  
%As FDI is closely related to strong-field double ionization (DI),

In this paper we numerically study FDI of atoms with circularly polarized light using a three-dimensional classical ensemble method. Our calculations reveal significant contributions of FDI and we show that its observation is more favourable for the targets with low ionization potentials and short laser wavelengths. We find that for high intensities where SDI dominates, the first ionized electron is more likely to be captured after the end of the laser pulse, which is in contrast with the case by linear polarization. For intensities corresponding to the  NSDI regime, the FDI probability as a function of intensity exhibits a ``knee'' structure. Furthermore, recollision plays an important role in FDI for all the intensities studied here. This can be experimentally verified by measuring the photoelectron momentum distribution from FDI.  
\section{Theoretical model}
Currently accurate quantum simulations of a two-active-electron system in a strong laser field still present a great challenge. Here we employ the well-established classical model \cite{panfili2001comparing} to study FDI of atoms with circularly polarized light. This model has shown remarkable success in explaining many important features of strong-field DI \cite{PhysRevLett.94.093002,PhysRevLett.101.113001,PhysRevLett.102.173002,PhysRevLett.110.073001,PhysRevA.92.033405,ben2017nonsequential,chen2016contribution}. Within this model, the evolution of a two-active-electron atom is described by the classical Newtonian equation of motion (atomic units are used throughout this paper):
\begin{equation}\label{Newton}
\frac{d^{2}\mathbf{%
r_{i}}}{dt^{2}}=-\mathbf{E}(t)-\nabla(V^{i}_{ne}+V_{ee}),
\end{equation}
where $\textbf{E}(t)=(E_{x}(t),E_{y}(t),0)$ is the circularly polarized laser field with $E_{x}(t)=\frac{E_{0}}{\sqrt{2}}f(t)\cos\omega t$ and $E_{y}(t)=\frac{E_{0}}{\sqrt{2}}f(t)\sin\omega t$ ($\omega$ is the laser frequency). Here $E_{0}$ is the peak amplitude of the laser electric field and $f(t)=\sin^{2}(\frac{\pi t}{10T})$ is the pulse envelope function with the full duration of 10$T$, where $T$ is the optical cycle. The index $i=1,2$ in Eq. (1) denote the two active electrons. The Coulomb interaction potential between the nucleus and the $i$th electron is $V^{i}_{ne}=-2/\sqrt{\mathbf{r}_{i}^{2}+a^{2}}$ and $V_{ee}=1/\sqrt{(\mathbf{r}_{1}-\mathbf{r}_{2})^{2}+b^{2}}$ represents the potential for the electron-electron interaction. The softening parameters $a$ and $b$ are introduced to avoid autoionization and numerical singularity \cite{PhysRevA.38.3430,PhysRevA.44.5997}. The values of $a$ and $b$ are set to be 3.0 a.u. and 0.05 a.u. for Mg atoms \cite{ben2017nonsequential}. For Ar atoms, we choose $a=1.5$ a.u. and $b=0.05$ a.u. \cite{2020WangxuOE}.   

The initial conditions of the two electrons are obtained by first randomly assigning their positions in the classical allowed region for the energy corresponding to the negative sum of the first and second ionization potentials of the target atom. The remaining energy is randomly distributed between the two electrons in momentum space. To obtain stable position and momentum distributions, the system is allowed to evolve without the laser field for a sufficiently long time (about 100 a.u.). The laser field is switched on once the initial ensemble is stable. The evolution of the two-electron system is traced until the end of the laser pulse according to the Newtonian equation. The energy of each electron
contains potential energy of the electron-ion interaction, the kinetic energy, and half of the electron-electron repulsion energy. A double-ionization event is identified when the final energies of both electrons are greater than zero. For FDI events, the energies of the two electrons achieve positive during the laser pulse \cite{2020WangxuOE,ourFDIpaper}. At the end of the laser pulse, one electron has positive final energy but the other is captured and has negative final energy above $-I_{p2}$, where $I_{p2}$ is the second ionization potential of the target atom.

\section{Numerical results and discussions}

\begin{figure}[h!]
\centering\includegraphics[width=7cm]{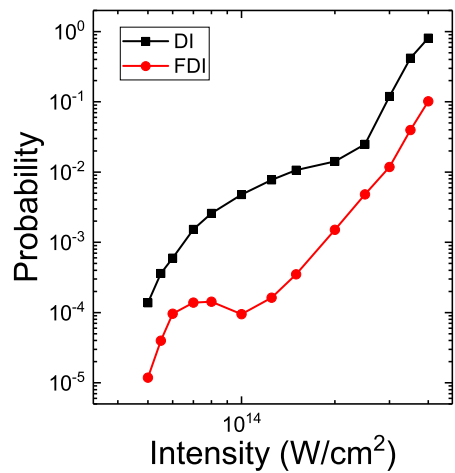}
\caption{Computed DI and FDI probabilities of Mg atoms as functions of intensity for 800-nm circularly polarized laser pulses.}
\end{figure}

Figure 1 shows the calculated probabilities of DI and FDI for Mg as functions of intensity. The characteristic ``knee'' structure indicating NSDI contributions below $\sim2\times10^{14}$ W/cm$^{2}$ can be clearly seen in DI results. Above this intensity, SDI starts to dominate. This is in good agreement with previous experiments \cite{PRA64.043413} and theoretical simulations \cite{PRL105.083001,PRL105.083002,PRL108.103601}. Our calculations also reveal the existence of significant FDI contributions by circular polarization. Interestingly, a similar ``knee'' shape below $1\times10^{14}$ W/cm$^{2}$ is also found in FDI results. Below we explore in detail the physical mechanism of FDI, which is the main focus of our paper. 

Our discussions start with the intensities where SDI dominates. In Fig. 2(a) we show the distribution of the closest distance between the two electrons after the emission of one electron from the core for FDI events at $3\times10^{14}$ W/cm$^{2}$. The ionization exit of the emitted electron ranges from 12 a.u. to 17 a.u. The first peak below 10 a.u. indicates that recollision occurs. This means that, surprisingly, for SDI regime where recollision plays a minor role, more than half FDI events are still connected to recollision. 
%recollision plays a minor role in double ionization at such high intensity. although double ionization mainly proceeds sequentially at such high intensity, more than half FDI events are still related to recollision.

To understand FDI dynamics, we first calculate ionization time distributions of the two electrons for the FDI events that are not related to recollision. Here the ionization time is defined when the energy of the electron becomes positive for the first time \cite{2020WangxuOE}. Within our model, the two electrons can be distinguished according to which one is captured at the end of the laser pulse. These two electrons are ionized independently and one cannot expect strong correlations between them. As seen in Fig. 2(b), the finally captured electron $e_1$ is ionized during the rising edge of the laser envelope, while the other electron $e_2$ (the photoelectron generated by FDI) is ionized around the peak of the laser envelope. This can be easily understood as the electron ionized around the peak obtains much larger drift momentum $-A(t_{e_2})$ from the laser field, making it more difficult to be captured at the end of the laser pulse. The calculations also demonstrate that for SDI regime, the early ionized electron tends to be captured, which is in striking contrast to the case by linearly polarized laser fields \cite{2020LarimianPRR}. A typical two-electron trajectory is displayed in Fig. 2(c). We find that the ionization-exit momentum of $e_1$ is not completely compensated by the drift momentum $-A(t_{e_1})$ obtained from the laser field (Fig. 3), revealing the important role of Coulomb focusing effects in FDI.

\begin{figure}[]
\centering\includegraphics[width=9cm]{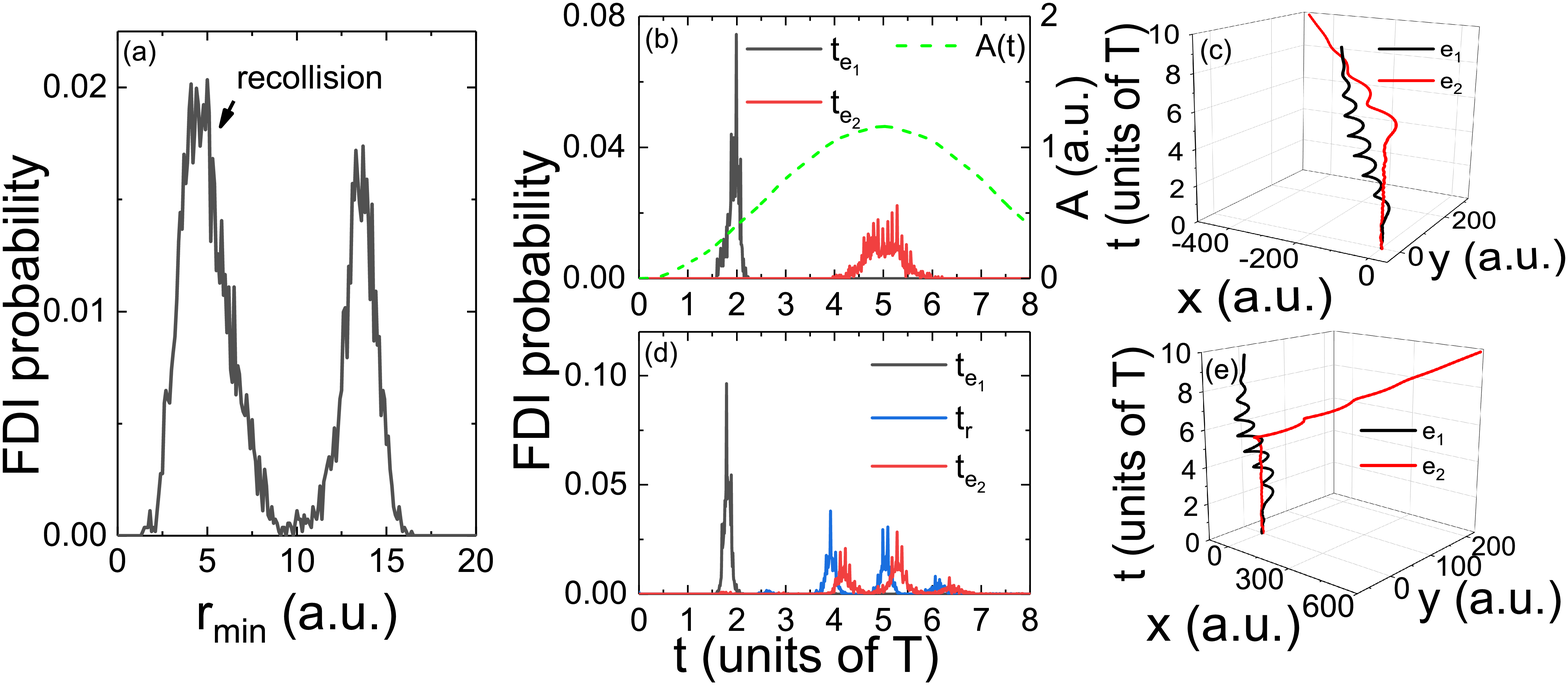}
\caption{(a) Calculated distribution of the shortest distance between the two electrons after departure of one electron from the core for FDI trajectories at $3\times10^{14}$ W/cm$^{2}$. (b) and (d) Ionization time distributions of the two electrons for FDI trajectories corresponding to the second and the first peak in (a), respectively. In (b) the absolute value of the vector potential as a function of time is shown. The recollision time distribution (blue line) is shown in (d). Typical time evolutions of the two-electron trajectories corresponding to the second and the first peak in (a) are shown in (c) and (e), respectively. Here $e_1$ and $e_2$ denote the finally captured electron and the photoelectron for FDI, respectively.}
\end{figure}

\begin{figure}[b!]
\centering\includegraphics[width=10cm]{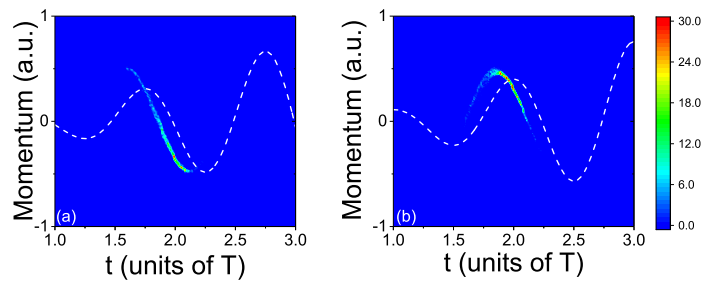}
\caption{Calculated probability distribution of ionization-exit momentum of $e_1$ along the $x$ (a) and $y$ (b) direction for the FDI events shown in Fig. 2(b). The corresponding vector potentials (white dotted lines) are also plotted.} 
\end{figure}

\begin{figure}[]
\centering\includegraphics[width=9cm]{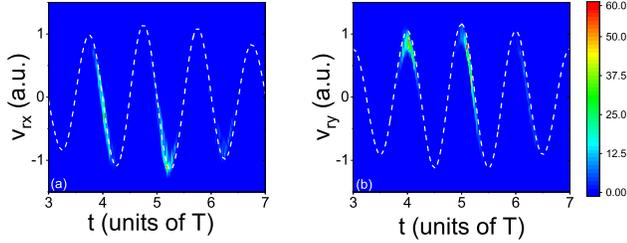}
\caption{(a) Calculated probability distribution of momentum of the recolliding electron at $t_{r}+0.1T$ along the $x$ direction. Here the FDI events are the same as those in Fig. 2(d). Also shown is the corresponding vector potential (white dotted line). (b) Same as (a) but for the $y$ direction.} 
\end{figure}

For the FDI events related to recollision, the dynamics is different. The finally captured electron is ionized during the rising edge and returns to the core several times around the peak of the laser envelope [Fig. 2(d)]. The calculated distribution of recollision time $t_{r}$ is shown in Fig. 2(d). Here the recollision time is defined as the instant of closest approach of the two electrons after departure of one electron from the core \cite{PhysRevLett.101.113001}. During recollision, a small portion of the energy of the returning electron is transferred to the second electron, which is ionized shortly after recollision and contributes to the photoelectron for FDI. In the polarization plane ($x-y$ plane), right after recollision both the residual momenta of the returning electron along the $x$ and $y$ axis, i.e., $v_{rx}$ and $v_{ry}$, are largely compensated by the drift momenta obtained from the laser field subsequently. This can be clearly seen in Fig. 4 where we show the calculated probability distributions of momenta of the returning electron at $t_{r}+0.1T$ and the corresponding vector potentials. Neglecting Coulomb potential, both the final velocities along the $x$ and $y$ directions $p_{x,y}\approx v_{rx,ry}-A_{x,y}(t_{r}+0.1T)$ are close to zero. Thus the rescattered electron tends to be captured after the end of the laser pulse. A typical two-electron trajectory showing such dynamics is depicted in Fig. 2(e).

\begin{figure}[b!]
\centering\includegraphics[width=9cm]{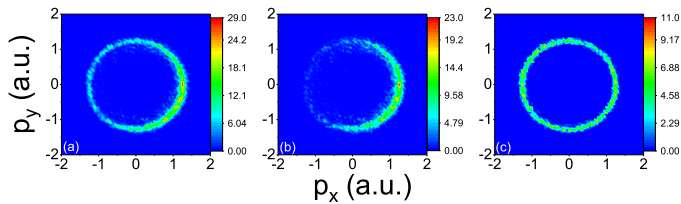}
\caption{(a) Calculated photoelectron momentum distribution for FDI at $3\times10^{14}$ W/cm$^{2}$. (b) and (c) Same as (a) but for the FDI events related and not related to recollision, respectively. }
\end{figure}

Next we show whether FDI is related to recollision can be inferred from the final momentum distribution of the photoelectrons produced by FDI, which can be measured using a reaction microscope. Figure. 5(a) shows the calculated photoelectron momentum distribution for FDI at $3\times10^{14}$ W/cm$^{2}$. One can find a doughnut-like structure with more electron yield in a crescent-shaped distribution. This is different from the electron momentum distribution from single ionization by circular polarization, which exhibits a symmetric doughnut-like shape \cite{PhysRevA.70.023413}. In Figs. 5(b) and 5(c) we show the electron momentum distributions for the FDI events related and not related to recollision, respectively. The comparison between Figs. 5(b) and 5(c) reveals that the asymmetric electron yields in Fig. 5(a) arise from recollision, which occurs every cycle mainly within the range from 3.5T to 5.5T [see Fig. 2(d)]. The ionization time distribution of the photoelectron shows similar features and each peak covers a half laser cycle [see Fig. 2(d)]. Therefore, the photoelectron obtains drift momentum from the laser field every half cycle, resulting in the crescent-shaped distribution shown in Fig. 5(b). For FDI trajectories not related to recollision, the ionization time distribution of the photoelectron almost covers one laser cycle [Fig. 2(b)], leading to the symmetric doughnut-like shape shown in Fig. 5(c).

\begin{figure}[]
\centering\includegraphics[width=9cm]{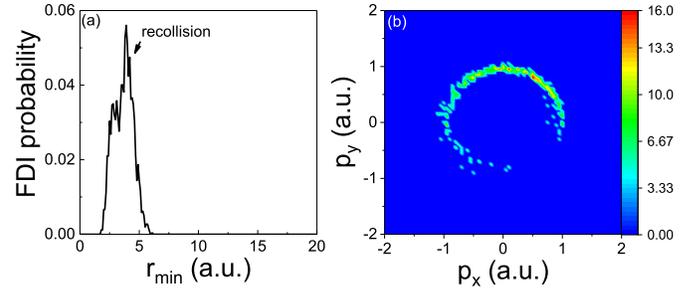}
\caption{(a) Same as Fig. 2(a) but for the FDI events at $1.5\times10^{14}$ W/cm$^{2}$. (b) Corresponding photoelectron momentum distribution for FDI events in (a). }
\end{figure}

\begin{figure}[b!]
\centering\includegraphics[width=8cm]{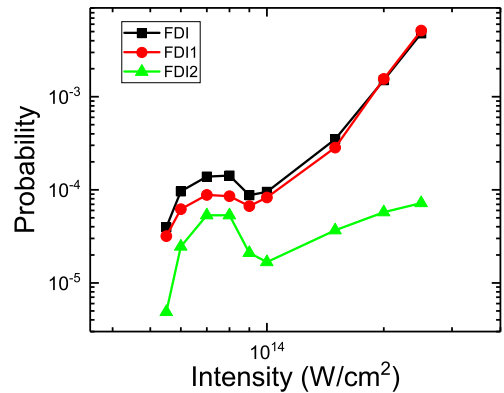}
\caption{Calculated probabilities of FDI, FDI1, and FDI2 for intensities corresponding to NSDI regime of Mg shown in Fig. 1.}
\end{figure}

For modest intensities where NSDI dominates, the mechanism of FDI is similar to that of the FDI events related to recollision at $3\times10^{14}$ W/cm$^{2}$. In Fig. 6(a) we show the calculated distribution of the closet approach of the two electrons after the ionization of one electron for FDI events at $1.5\times10^{14}$ W/cm$^{2}$. The calculation shows a single peak below 10 a.u., indicating that all FDI events are related to recollision. The ionization of the photoelectron is closely connected to the recollision, which occurs every cycle around the peak of the laser envelope (not shown). As explained above, this leads to the crescent-shaped distribution of the photoelectron momentum distribution shown in Fig. 6(b). Again, the calculations indicate that recollision leaves its footprints in the photoelectron momentum distribution from FDI with circular polarization, which can be experimentally verified. 
%This can be used to identify recollision effects for circularly polarized light. 

Another interesting feature of FDI in Fig. 1 is that the dependence of its probability on intensity exhibits a ``knee'' structure appearing at intensities below $1\times10^{14}$ W/cm$^{2}$. To understand this feature, we separate the FDI events related to recollision into FDI1 and FDI2 events. Here FDI1 and FDI2 correspond to that the recolliding electron and the other electron are captured after the laser turnoff, respectively. Figure. 6 shows the calculated probabilities of FDI1 and FDI2 for the intensities where NSDI dominates. One can find that the ``knee'' shape of FDI arises from the distinct intensity dependences of FDI1 and FDI2. Above $1\times10^{14}$ W/cm$^{2}$, FDI1 events constitute the main contribution to FDI events. For intensities around $7.5\times10^{13}$ W/cm$^{2}$, FDI2 contributions also play a significant role. We have found that for such intensities the doubly excited states (DESs) of Mg are largely populated shortly after recollision, which is similar to the previous study using linearly polarized light \cite{ourFDIpaper}. From the recollision-induced DESs, the two electrons experience almost the same laser electric field afterwards and there is no preference for each electron with higher final energy than the other. This leads to the comparable contributions from FDI1 and FDI2 \cite{ourFDIpaper}. 
Due to the similar dynamics of FDI1 (FDI2) around $7.5\times10^{13}$ W/cm$^{2}$, i.e., the DESs serve as the main pathway leading to FDI1 (FDI2), both the intensity dependences of the probabilities of FDI1 and FDI2 become flat. 
%Around $7.5\times10^{13}$ W/cm$^{2}$ the DESs serve as the main pathway leading to FDI1 and also FDI2. Due to the similar dynamics for those intensities, both the intensity dependences of the probabilities of FDI1 and FDI2 become flat.  
%As the DESs serve as the main pathway leading to FDI1 and also FDI2, both the intensity dependences of the probabilities of FDI1 and FDI2 become flat around $7.5\times10^{13}$ W/cm$^{2}$.          

\begin{figure}[]
\centering\includegraphics[width=9cm]{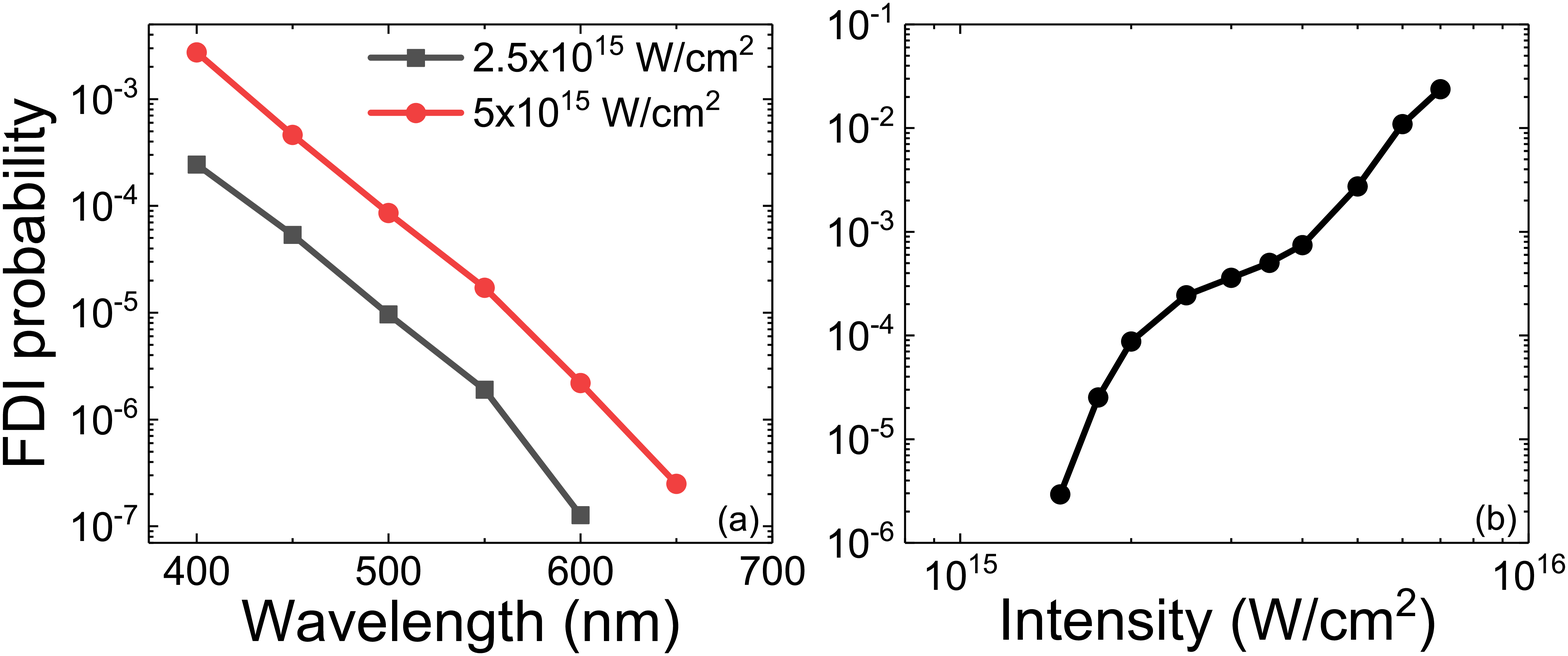}
\caption{(a) Calculated FDI probabilities of Ar atoms as functions of wavelength at the intensities of $2.5\times10^{15}$ W/cm$^{2}$ and $5\times10^{15}$ W/cm$^{2}$ for circular polarization. (b) Calculated FDI probability of Ar as a function of intensity for 400 nm circular polarization. Other laser parameters are the same as Fig. 1.}
\end{figure}

It is well known that the ``knee'' structure of DI probability as a function of intensity is a consequence of recollision. Furthermore, it has been shown that recollision is universal in circularly polarized laser fields \cite{PhysRevA.95.013402} and to observe this ``knee'' structure the laser frequency has to be larger than $0.18(I_{p1})^{5/4}$, where $I_{p1}$ is the first ionization potential of the target \cite{PRL108.103601}. Therefore, the observation of the ``knee'' shape by circular polarization favours targets with low ionization potentials such as Mg and short driving laser wavelengths. This is why the ``knee'' structure has escaped observation for DI of rare gas atoms with circular polarization at 800 nm. For FDI, we have demonstrated the crucial role of recollision for the whole range of intensities covering both NSDI and SDI regimes. Consequently, one can expect that FDI probability will be significantly decreased for atoms with high ionization potentials and for long wavelengths where recollision is suppressed. In Fig. 8(a) we show the calculated FDI probabilities of Ar atoms as functions of wavelength for two different intensities. Indeed, the FDI probability decreases rapidly as the wavelength is increased. For wavelengths longer than 650 nm, the FDI probability is too small to be calculated, which is in accordance with the absence of NSDI of Ar for such wavelengths \cite{PhysRevA.95.013402,PRL108.103601}. Figure. 8(b) shows the FDI probability of Ar as a function of intensity for 400 nm circularly polarized laser fields. One can see a ``knee'' structure similar to the FDI calculation for Mg at 800 nm shown in Fig. 1. This is in line with the prediction for DI that recollision is more favourable for rare gases with short wavelengths \cite{PhysRevA.95.013402}.

%It is well known that the ``knee'' structure of DI probability as a function of intensity is a consequence of recollision. Differently, the similar ``knee'' structure revealed in our FDI calculation is caused by the different dependences of FDI1 and FDI2 on intensity, which arises from the fact that the recollision-induced DESs serve as the main pathway leading to FDI for low intensities. For relatively high intensities corresponding to NSDI regime,      
%although NSDI starts to dominate below $\sim2\times10^{14}$ W/cm$^{2}$, the FDI probability first decreases rapidly and then shows a knee structure below $1\times10^{14}$ W/cm$^{2}$ as the intensity is decreased. 

\section{Conclusion}
In conclusion, we have theoretically investigated FDI of atoms exposed to intense circularly polarized laser pulses. The simulations reveal a novel ``knee'' structure of FDI probability as a function of intensity, which is similar to DI results at modest intensities. We demonstrate that, to observe FDI with circular polarization, it is beneficial to employ targets with low ionization potentials and short laser wavelengths. This is due to the fact that recollision plays an essential role in FDI not only for the NSDI regime but also for the SDI regime. We further show that recollision can be experimentally identified from the photoelectron momentum distribution produced by FDI. 

\section*{ACKNOWLEDGMENTS}
This work was supported the National Natural Science Foundation of China (11974380), Deutsche Forschungsgemeinschaft (DFG) in the framework of the Schwerpunktprogramm (SPP) 1840, and Quantum Dynamics in Tailored Intense Fields (QUTIF).

\bibliographystyle{apsrev4-1} % Tell bibtex which bibliography style to use
\bibliography{FDI} % Tell bibtex which .bib file to use (this one is some example file in TexLive's file tree)
\end{document}